\documentclass{ws-ijmpa}
\usepackage[super,compress]{cite}
\usepackage{multirow}
\usepackage{amssymb}
\usepackage{amsmath, graphicx}
\usepackage{dcolumn}
\usepackage{bm}
\usepackage{enumerate}
\usepackage{slashed}
\usepackage{epstopdf}
\usepackage[unicode]{hyperref}
\usepackage{color}
\begin{document}
\markboth{Vo Quoc Phong and Nguyen Minh Anh}{Electro-weak Phase Transition With Three Phases in The $SU(2)_1 \otimes SU(2)_2 \otimes U(1)_Y$ Model}

%
\catchline{}{}{}{}{}
%

\title{Electro-weak Phase Transition With Three Phases in The $SU(2)_1 \otimes SU(2)_2 \otimes U(1)_Y$ Model}

\author{Vo Quoc Phong}
\address{Department of Theoretical Physics, VNUHCM-University of Science,\\
Ho Chi Minh City, Vietnam}
\author{Nguyen Minh Anh}
\address{Department of Theoretical Physics, VNUHCM-University of Science,\\
	Ho Chi Minh City, Vietnam}
\maketitle

\begin{history}
\received{Day Month Year}
\revised{Day Month Year}
\end{history}

\begin{abstract}
	Our analysis shows that SM-like electroweak phase transition (EWPT) in the $SU(2)_1 \otimes SU(2)_2 \otimes U(1)_Y$ (2-2-1) model is a first-order phase transition at the $200$ GeV scale (the SM scale). Its strength ($S$) is about $1 - 2.7$ and the masses of new gauge bosons are larger than $1.7$ TeV when the second VEV is larger than $535$ GeV in a three-stage EWPT scenario and the coupling constant of $SU(2)_2$ group must be larger than 2. Therefore, this first order EWPT can be used to fix VEVs and the coupling constant of the gauge group in electro-weak models.
\end{abstract}

\ccode{PACS numbers: 11.15.Ex, 12.60.Fr, 98.80.Cq}

\section{INTRODUCTION}\label{secInt}

The Electro-weak Baryogenesis (EWBG) is a way to explain the Baryon Asymmetry of the Universe (BAU), has been known by Three Sakharov conditions Ref.~[\citen{sakharov}]. The third condition is a first-order EWPT, not only leads to thermal imbalance Ref.~[\citen{mkn}] but also makes a connection between B and CP violation via non-equilibrium physics Ref.~[\citen{ckn}]. The B violations can be showed throughout the sphaleron rate which must satisfy the decoupling conditions Ref.~[\citen{sphaleron,sphaleronB,sphaleronC,sphaleronD,sphaleronE,sphaleronF,sphaleronG}].

The toy hight-temperature effective potential, at the one-loop level is usually calculated as follows:
\[ V_{eff}=D.(T^2-{T}^2_0){v}^2-E. Tv^3+\frac{\lambda_T}{4}v^4, \]
where $v$ is the VEV of Higgs. This potential has two minima at $T<T_C$ ($T_C$ is a critical temperature). The depth of the second minimum at $T_C$, $v_C=\frac{2E.T_C}{\lambda_{T_C}}$. We can see a simple thing that the effective potential has not a second minimum when $E=0$, we have a second-order phase transition.

The true act of the decoupling condition, the sphaleron rate is smaller than the cosmological expansion rate at temperatures below the critical temperature; i.e., $\frac{v_c}{T_C}$ which can be called the EWPT strength, must be larger than unity Ref.~[\citen{sphaleron}].

The EWPT has been investigated in the Standard Model (SM) Ref.~[\citen{mkn,SME,michela}] as well as beyond SM Refs.~[\citen{BSM,BSMB,BSMC,majorana,thdm,ESMCO,elptdm,elptdmB,elptdmC,elptdmD,elptdmE,phonglongvan,phonglongvanB,phonglongvanC,SMS,dssm,munusm,lr,singlet,singletB,singletC,singletD,mssm1,mssm1B,twostep,twostepB,twostepC,1101.4665,1101.4665B,1101.4665C}]. The EWPT strength is larger than unity at the 200 GeV scale in SM, but the mass of Higgs boson must be less than $125$ GeV Refs.~[\citen{mkn,SME,SMEB,SMEC,michela}].

Many extensions have triggers for the first-order EWPT, heavy bosons or dark matter candidates  Refs.~[\citen{majorana,thdm,thdmB,ESMCO,elptdm,elptdmB,elptdmC,elptdmD,elptdmE,phonglongvan,phonglongvanB,phonglongvanC,epjc,zb,singlet,singletB,singletC,singletD,mssm1,mssm1B,twostep,twostepB,twostepC,chiang3}]. Another pretty important point is that there are proofs that EWPT does not depend on the gauge. This allows us to calculate EWPT in the Landau gauge as simplest and also physically adequate Ref.~[\citen{zb,1101.4665,1101.4665B,1101.4665C,Arefe}]. In models with more doubly charged particles or bosons, the strength will be larger Ref.~[\citen{chiang3}].

The $SU(2)_1 \otimes SU(2)_2 \otimes U(1)_Y$ Model (2-2-1 model) is one extensions of SM, which has a group structure close to SM. However, there are three coupling constants, three VEVs; two exotic quarks which are in a doublet of $SU(2)_2$ group; one new charged and one new neutral gauge boson which are larger than $1.7$ TeV Ref.~[\citen{221}]. This model has two new gauge bosons which can play an important role in the early universe. These particles and the frame of Higgs potential can be a reason for one first-order EWPT.

This article is organized as follows. In Sect.\ref{sec2}, a short review of the 2-2-1 model and the corresponding Higgs potential will be presented. The electroweak phase transition structure will be driven in Sect.\ref{sec3}. The range of mass of charged scalar particles and the coupling constant of $SU(2)_1$ group are found by a first-order phase transition condition in Sect.\ref{sec4}. Finally, in Sect.\ref{sec8} we summarize and describe outlooks for this work.

\section{Review on 2-2-1 model}\label{sec2}
In this model Ref.~[\citen{221}], Fermion sectors are like SM,
\begin{align}
\begin{pmatrix}
u_L \\
d_L
\end{pmatrix}
;
\begin{pmatrix}
c_L \\
s_L
\end{pmatrix}
;
\begin{pmatrix}
t_L \\
b_L
\end{pmatrix}
,
\nonumber
\quad \begin{pmatrix}
\nu _{eL} \\
e_L
\end{pmatrix}
;
\begin{pmatrix}
\nu _{\mu L} \\
\mu _L
\end{pmatrix}
;
\begin{pmatrix}
\nu _{\tau L} \\
\tau _L
\end{pmatrix}
,
\nonumber
\\
u_R; c _R; t _R,
\nonumber
\quad d_R; s _R; b _R,
\nonumber
\quad e_R; \mu _R; \tau _R.\nonumber
\end{align}
The electric charges of particles are defined $Q_{em}= T_3^{(1)} + T_3^{(2)} +Y$, with $T_3^{(1,2)} = \dfrac{\sigma_3}{2} $ and $\sigma_3$ is the third Pauli matrix. The SM particles can be found in the representations of $SU(2)_1\otimes U(1)_Y$ and singlets of $SU(2)_2$. Besides, a vector-like quark doublet (VLQ) $Q'^T=(U',D')$, with the left-handed and right-handed chiralities transform in the same way as in Table \ref{rep}, is introduced Ref.~[\citen{221}]. They minimize the number of particles and increase the decays of heavy scalar boson or Higgs. Because the VLQs loops can be contribute to the decay of heavy Higss and the new leptons are not necessary to cancel the gauge anomaly Ref.~[\citen{221}].

\begin{table}[!ht]
\centering
\tbl{Representations and charge assignments of particles}
{\begin{tabular}{||c||ccccccccc||}			
\hline 
\hline
&\multicolumn{6}{|c|}{Fermions} & \multicolumn{3}{|c||}{Scalar}\\
\hline
&$Q_L$ & $u_R$ & $d_R$ & $L_L$ & $e_R$ & $Q'_{L(R)}$ & \multicolumn{3}{|c||}{$H_1$  \quad $H_2$ \quad $S'$}\\
\hline 
$SU_C(3)$ &$\textbf{3}$ &$\textbf{3}$& $\textbf{3}$& $\textbf{1}$& $\textbf{1}$& $\textbf{3}$& $\textbf{1}$& $\textbf{1}$& $\textbf{1}$   \\ 
\hline
$SU(2)_1$ &$\textbf{2}$ &$\textbf{1}$& $\textbf{1}$& $\textbf{2}$& $\textbf{1}$& $\textbf{1}$& $\textbf{2}$& $\textbf{1}$& $\textbf{1}$  \\
\hline
$SU(2)_2$ &$\textbf{1}$ &$\textbf{1}$& $\textbf{1}$& $\textbf{1}$& $\textbf{1}$& $\textbf{2}$& $\textbf{1}$& $\textbf{2}$& $\textbf{1}$  \\
\hline
$U(1)_Y$ & 1/6 & 2/3 & -1/3 & -1/2 &-1& 1/6& 1/2& 1/2& 0\\
\hline
\hline
\end{tabular}}\label{rep}	
\end{table}
        
\subsection{Higgs potential}
The Higgs potential with two doublets and one singlet, is given by:
\begin{align}
\nonumber
V(H_1,H_2,S') = &\sum_{i=1,2}[\mu_1^2 H^\dagger_i H_i + \lambda_i (H^\dagger_i H_i)^2] + \mu_s^2 S'^2 + \lambda_S S'^4+\mu_3 S'^3 \\
&+\lambda_{12}H^\dagger_1 H_1 H^\dagger_2 H_2 +\lambda_{1S} S'^2 H^\dagger_1 H_1+\lambda_{2S}S'^2 H^\dagger_2 H_2\\\nonumber
&+ S'(\mu_{1S} H^\dagger_1 H_1 + \mu_{2S} H^\dagger_2 H_2).
\end{align}

The scalar fields in $V(H_1,H_2,S')$ can be expressed as:
\begin{align}
H_i= 
\begin{pmatrix}
G_i^\dagger\\
(v_i + h_i + iG_i^0)/\sqrt{2}
\end{pmatrix},\qquad S'=(v_S+S)/\sqrt{2},
\end{align}
where $G^+_i,G^0_i$ are the Nambu-Goldstone bosons; $h_{1,2} $ and $S$ are the scalar bosons; $v_{1,2,S}$ are the vevs of Higgs fields. The $S$ field directly couples to the heavy VLQs. S plays the role of mass generating for VLQs. They do not directly couple to SM-particles so they are candidates for dark matter. 

From the above potential, we have the mass-square matrix for the scalar bosons Ref.~[\citen{221}]: 
\begin{align*}
\mathcal{M}^2=\left(\begin{matrix}
m^2_{h_1}&\lambda_{12}v_1v_2&\lambda_{1S}v_1v_S+\frac{\mu_{1S}v_1}{\sqrt{2}}\\
\lambda_{12}v_1v_2&m^2_{h_2}&\lambda_{2S}v_2v_S+\frac{\mu_{2S}v_2}{\sqrt{2}}\\
\lambda_{1S}v_1v_S+\frac{\mu_{1S}v_1}{\sqrt{2}}&\lambda_{2S}v_2v_S+\frac{\mu_{2S}v_2}{\sqrt{2}}&m^2_{S}
\end{matrix}\right), 
\end{align*}

where the masses of Higgs bosons Ref.~[\citen{221}] are:
\begin{align}
\begin{split}
&m^2_h= m^2_{h_1}= 2\lambda_1 v_1^2, \qquad m^2_{h_2}= 2\lambda_2 v_2^2\\
&m_S^2=2\lambda_S v_S^2 + \frac{3\mu_S v_S}{2\sqrt{2}}-\frac{\mu_{1S} v^2 +\mu_{2S} v_2^2}{2\sqrt{2} v_S}.
\end{split}
\end{align}

The fields $h_2, S$ are not physical states. So it will lead to an introduction of a $\phi$ mixing angle which $\sin 2\phi=2m_{23}^2/(m^2_{H_S}-m^2_H)$. Therefore, we obtain two physical particles and their masses Ref.~[\citen{221}] are
\begin{equation}
m^2_{H/H_S}=\frac{m^2_{S}+m^2_{h_2}}{2}\pm \frac{1}{2}\sqrt{(m^2_{S}-m^2_{h_2})^2+4m^4_{23}},
\end{equation}
with $m^2_{23}=\lambda_{2S}v_2v_S+v_2\mu_{2S}/\sqrt{2}$. However, we approximate that $\mu_{2S}$ and $\lambda_{2S}$ are very small Ref.~[\citen{221}]. The parameters $\lambda_{12}, \lambda_{1S}$ are the coupling constants of $h_1-h_2$ and $h_1-S$. They must be small so that the Higgs decays are not too large Ref.~[\citen{221}]. It is clear that the off-diagonal elements of $\mathcal{M}^2$ will be small. Therefore $m_{23} \sim 0$, $m_{H_S}\approx m_S, m_{H}\approx m_{h_2}$. The particle $h_1$ is considered as the SM-like Higgs $h$ so we use $h,v$ instead of $h_1,v_1$ from now on.

\subsection{Gauge boson sector}
The masses of the gauge bosons can be found in the kinetic part of the Lagrangian
\begin{align}
\mathcal{L}= (D_\mu H_1)^\dagger (D_\mu H_1) + (D_\mu H_2)^\dagger (D_\mu H_2)+ (D_\mu S')^\dagger (D_\mu S').
\end{align}

We can find the masses of gauge bosons by writing the convariant derivative as:
\begin{align}
D_\mu=(\partial_\mu - ig_i T_a^{(i)} A_{i\mu}^a - ig_Y YB_\mu ),
\end{align}
where $g_i$ and $ A_{i\mu}^a$ $(a=1,2,3)$ are the gauge coupling and gauge fields of $SU(2)_i$, $g_Y$ and $B_\mu $ are the gauge coupling ang gauge field of $U(1)_Y$, $T_a^{(i)}=\sigma_a/2$, where $\sigma_a$ are the Pauli matrices and $Y$ is the hypercharge of a particle. We can easily obtain the masses of SM-like and a new charged gauge boson as:
\begin{align*}
m_W=\dfrac{g_1v}{2} \qquad \text{and} \qquad m_{W'}= \dfrac{g_2 v_2}{2}.
\end{align*}

The physical masses of the two neutral gauge bosons $Z$ and $Z'$ are:
\begin{align*}
m^2_{Z/Z'} = \dfrac{m^2_{Z_1}+m^2_{Z_2}}{2}\pm\dfrac{1}{2} \sqrt{(m^2_{Z_2}-m^2_{Z_1})^2+ 4m^4_{Z_1 Z_2}},
\end{align*}
where
\begin{align*}
m^2_{Z_1}= \dfrac{v^2}{4} (g_1^2+g'^2),\quad m^2_{Z_2}= \dfrac{v_2^2 g_2^4 + v^2 g'^4}{4(g_2^2- g'^2)},\quad m^2_{Z_1 Z_2}= \dfrac{v^2 g'^2}{4} \sqrt{\dfrac{g_1^2+ g'^2}{g_2^2- g'^2}},
\end{align*}
with
\begin{align*}
g'=g_Y\cos{\theta}, \cos{\theta}=\frac{g_2}{\sqrt{g_2^2+g^2_Y}}.
\end{align*}

Finally, the Yukawa sector as follows:
\begin{align}
-\mathcal{L}= y_F \bar{Q}'_L Q'_R S' + y_b \bar{Q}'_L H_2 b_R +y_t \bar{Q}'_L \tilde{H_2}  t_R + m_\psi \bar{Q}'_L Q'_R + H.c
\end{align}
\section{Electroweak phase transition structure in the 2-2-1 model}\label{sec3}
The purpose of this section is to find the effective potential of 2-2-1 model. The process will be similar to the one of SM. Higgs components and gauge bosons are main contributors to EWPT, so determining the mass of these particles can affect the phase separation.

First, we have the Higgs Lagrangian of 2-2-1 model, which contains the kinetic energy and potential parts as:
\begin{align}
\mathcal{L}_\text{Higgs}= (D_\mu H_1)^\dagger (D_\mu H_1) + (D_\mu H_2)^\dagger (D_\mu H_2)+ (D_\mu S')^\dagger (D_\mu S')+ V(H_1,H_2,S').
\end{align}

After averaging over all space, we get:
\begin{align}
&\langle H_i \rangle = \dfrac{1}{\sqrt{2}}
\begin{pmatrix}
0\\
v_i
\end{pmatrix}, \\
&\langle S'\rangle= \dfrac{1}{\sqrt{2}} v_S; \qquad i=1,2.
\end{align}

Lagrangian is rewritten as below since we can consider $v,v_2$ and $v_S$ as variables from now on.
\begin{align}
\mathcal{L}_{Higgs}=& \dfrac{1}{2} \partial^\mu v \partial_\mu v + \dfrac{1}{2} \partial^\mu v_2 \partial_\mu v_2 + \dfrac{1}{2} \partial^\mu v_S \partial_\mu v_S + V_0(v,v,v_S)\nonumber\\
&+\sum_{i =\text{vector boson}}m_i^2(v,v_2,v_S)W^\mu W_\mu+\sum_{j =\text{scalar boson}}m_j^2(v,v_2,v_S)H^2, 
\end{align}
in which $W$ and $H$ run over all vector and scalar boson, respectively.
	
\begin{table}[htbp]	
	\tbl{Masses of bosons and fermions in 2-2-1 model.}
{\begin{tabular}{|c|c|c|c|c|c|} 
	\hline
	Particles & $m^2(v,v_2,v_S)$ & $m^2(v)$ & $m^2(v_2)$& $m^2(v_S)$ & $n$ \\ 
	\hline
	$m^2_{W^\pm}$ & $\frac{g^2v^2}{4}$ &$\frac{g^2v^2}{4}$& 0 & 0 & $6 $\\ 
	
	$m^2_{W'^\pm}$ & $\frac{g_2^2 v_2^2}{4}$ &0 & $\frac{g_2^2 v_2^2}{4}$ &0& $6$ \\ 
	
	$m^2_{Z_1} \sim m^2_Z$ & $(g^2+ g'^2)\frac{v^2}{4}$ & $(g^2+ g'^2)\frac{v^2}{4}$ &0& 0& $3$ \\
	
	$m^2_{Z_2} \sim m^2_{Z'}$ & $\frac14\frac{g'^4 v^2 + g_2^4 v_2^2}{g_2^2 - g'^2}$ & $\frac14\frac{g'^4 v^2}{g_2^2 - g'^2}$ & $\frac14\frac{g_2^4 v_2^2}{g_2^2 - g'^2}$& 0& $3$ \\
	
	$m^2_h= m^2_{h_1}$ & $ 2\lambda_1 v^2$ &$ 2\lambda_1 v^2$&0&0& $1$ \\
	
	$m^2_H= m^2_{h_2}$ & $ 2\lambda_2 v_2^2$&0& $ 2\lambda_2 v_2^2$&0 & $1$ \\
	
	$m^2_{H_S}=m_S^2$ & $2\lambda_S v_S^2 + \frac{3\mu_S v_S}{2\sqrt{2}}-\frac{\mu_{1S} v^2 +\mu_{2S} v_2^2}{2\sqrt{2} v_S}$ & $-\frac{\mu_{1S} v^2}{2\sqrt{2} v_S}$& $-\frac{\mu_{2S} v_2^2}{2\sqrt{2} v_S}$& $2\lambda_S v_S^2 + \frac{3\mu_S v_S}{2\sqrt{2}}$& $1$ \\
	
	$m_t^2$ & $f_t^2 v^2$ &$f_t^2 v^2$&0&0& $-12$ \\
	
	$m_T^2 \sim m_{U'}^2 = m^2_Q$ & $(m_\psi+\frac{y_F}{\sqrt{2}}v_S)^2$ &0&0& $(m_\psi+\frac{y_F}{\sqrt{2}}v_S)^2$& $-12$ \\
	$m_B^2 \sim m_{D'}^2 = m^2_Q$ & $(m_\psi+\frac{y_F}{\sqrt{2}}v_S)^2$ &0&0& $(m_\psi+\frac{y_F}{\sqrt{2}}v_S)^2$& $-12$ \\
	\hline
\end{tabular}}\label{masss}
\end{table}

Table \ref{masss} contains the masses of the particles in this model Ref.~[\citen{221}], which depend on the VEVs; $n$ is the degree of freedom of the particles; $g_1=0.654, g'=0.407$; $g_2$ is unknown and it should be larger than 2 Ref.~[\citen{221}]. We can split the masses of particles into 3 parts as:
\begin{align}
m^2(v_S,v_2,v)= m^2(v_S)+m^2(v_2)+m^2(v).
\end{align}
The tree potential $V_0$ has the form:
\begin{align}\label{the}
V_0(v,v_2,v_S)&= V(\langle H_1\rangle,\langle H_2 \rangle,\langle S'\rangle)\\\nonumber
&= \dfrac{\mu_1^2}{2} v^2 + \dfrac{\mu_2^2}{2} v_2^2+ \dfrac{\lambda_1}{4} v^4 + \dfrac{\lambda_2}{4} v_2^4 + \dfrac{\mu_S^2}{2} v_S^2 + \dfrac{\lambda_S}{4} v_S^4 \\\nonumber
& + \dfrac{\mu_3}{2\sqrt{2}}v_S^3+ \dfrac{1}{2\sqrt{2}} v_S (\mu_{1S} v^2 + \mu_{2S} v_2^2)+ \lambda_{12} v^2 v_2^2 + \lambda_{1S} v_S^2 v^2 + \lambda_{2S} v_S^2 v_2^2\\
&=V_0(v_S)+V_0(v_2)+V_0(v)+\text{mixing terms}.
\end{align}

If we do not neglect these mixing terms, $V_0$ will have additional components $v^2_2 v^2$, $v^2_Sv^2$, $v^2_2v^2_S$. In general, at a temperature T, the effective potential depend on VEV $v$, will be a example form:

\begin{equation}
V_ {eff} (v) = \Lambda v^4-Ev^3 + \mathcal{D}v^2 + \lambda_k. v_2^2 v^2 + \lambda_j. v_S^2 v^2\approx \Lambda v^4-Ev^3 + \mathcal{D}v^2 + \lambda_i. (v^2_S + v^2_2) v^2,
\end{equation} 
where $\lambda_i$ represents $\mu_{1S},\lambda_{1S}, \mu_{1S}$ or $\lambda_{12}$. $\lambda_{12}$ and $\lambda_{1S}$ must be small unless they may cause a too large Higgs production cross section and BR for the Higgs to diphoton decay Ref.~[\citen{221}]. 
	
At a slice with $(v^2_S+v^2_2)$, $V_ {eff} (v)$ has two minima. But when the larger $\lambda_i$ is, the clearer second minimum is not. In other works, when $\lambda_i$ is very lager, it breaks the two-minimum structure of the effective potential. Because of a strong EWPT, $\lambda_i$ must be enough small and the effective potential remains the same in an absence of $\lambda_i$. Furthermore, the high order corrections of Higgs decays should not be divergent, $\lambda_i$ must be small. Therefore the mixing term in Eq.~\ref{the} can be neglect or it has little effect on EWPT. This is also detailed in Ref.~[\citen{ptl}].

In the scenario for the symmetry breaking in 2-2-1 model, there are 3 phase transitions at 3 different scales. The first with the scale $v_{S0}\sim \text{few TeV}$, generates the masses for exotic quarks. After that, the symmetry breaking  $SU(2)_1\otimes SU(2)_2\otimes U(1)_Y \rightarrow SU(2)_L\otimes U(1)_Y$ continues the job and generates the masses for two new heavy gauge bosons $W'$ and one part of $Z'$ through $v_2$. Finally, when the universe cools down to the electroweak scale $246$ GeV, the last symmetry breaking $SU(2)_L\otimes U(1)_Y \rightarrow U(1)_Q$ through $v$, generates the masses for the SM-like particles and the other part of $Z'$ boson. 

Multi-stage EWPT has been considered in many beyond SM models. Separation into several phases is due to the square of particle mass without the mixing of VEVs (except $H_S$). This problem may be well addressed in Ref.~[\citen{epjc}]. 

The mass of $H_S$ has a mixing of VEVs because the Higgs potential has the interaction among $S'$, $H_1$ and $H_2$, $S'(\mu_{1S} H^\dagger_1 H_1 + \mu_{2S} H^\dagger_2 H_2)$. This will lead to a difficulty in separating. In the next section we will approximate the mass of $H_S$, it can participate in one or two phases.

\section{Electroweak Phase Transition}\label{sec4}

\subsection{Mass generation for vector-like Quark $Q'$}
This phase involves one heavy Higgs boson $H_S$, two exotic quarks, and no SM particles. This phase transition is just a mediate stage where the exotic quarks can get their masses by interacting with the heavy Higgs field $H_S$. This process is purely Yukawa interaction.

\subsection{2nd phase transition $SU(2)_1 \otimes SU(2)_2\otimes U(1)_Y \rightarrow SU(2)_L\otimes U(1)_Y$}
This phase transition involves a partly mass of new gauge bosons $W'^\pm,Z'$, a partly mass of two new Higgs bosons $H,H_S$. The masses of them are functions of $v_2$ as the 4th column in table \ref{masss}. 

In Table \ref{masss}, the partly mass of $H_S$ in this phase is $-\dfrac{\mu_{2S} v_2^2}{2\sqrt{2} v_S}$ which depend on $v_2, v_S$. With $v_2\ll v_S$ and the dynamics variable is $v_2$ so we can approximate $-\dfrac{\mu_{2S}}{2\sqrt{2} v_S}\sim const$, i.e., we consider the contribution of $H_S$ with an effective mass $m_{H_S}^2(v_2)\sim const.v^2_{2}$. 

The one-loop effective potential includes the 0K and thermal contribution where the
former is $V^{0K}_{eff}(v_2)$, the latter is $V^{T}_{eff}(v_2)$. $V^{0K}_{eff}(v_2)$ include the tree potential and the vacuum contribution with the one loop corrections of all particles. After summing all one-loop diagrams, we obtain the vacuum contribution as follows:  

\begin{align*}
V^{0K}_{eff}(v_2) =& V_0(v_2)+ \dfrac{1}{64\pi^2}\bigg(6 m^4_{W'}(v_2)\ln\dfrac{m^2_{W'}(v_2)}{Q^2}+ 3m^4_{Z'}(v_2)\ln\dfrac{m^2_{Z'}(v_2)}{Q^2}\\
&+m^4_{H}(v_2)\ln\dfrac{m^2_{H}(v_2)}{Q^2}+m^4_{H_S}(v_2)\ln\dfrac{m^2_{H_S}(v_2)}{Q^2} \bigg).
\end{align*}

Using the Bose-Enstein and Fermi-Dirac distribution for boson and fermion, respectively, we obtain the following expression for the thermal contribution to $W', Z', H, H_S$:
\begin{align*}
V^{T}_{eff}(v_2) =\dfrac{T^4}{4\pi^2} \bigg[ 6F_-\left( \dfrac{m_{W'}(v_2)}{T} \right)+ 3F_-\left( \dfrac{m_{Z'}(v_2)}{T} \right) +F_-\left( \dfrac{m_{H}(v_2)}{T} \right)+ F_-\left( \dfrac{m_{H_S}(v_2)}{T} \right) \bigg],
\end{align*}
where
\begin{align}
&F_\pm\left(\dfrac{m_\phi}{T}\right)=\int_0^{\frac{m_\phi}{T}} \alpha J_\pm^{(1)}(\alpha,0)d\alpha, \quad J_{\pm}^{(1)}(\alpha,0)= 2\int_\alpha^{\infty} \dfrac{(x^2-\alpha^2)^{\nu/2}}{e^{x} \pm 1}dx \label{ac1}\\
\Rightarrow
&\begin{cases}
&J_{-}^{(1)}(\alpha,0)= \dfrac{\pi^2}{3}-\pi\alpha- \dfrac{\alpha^2}{2}(\ln\dfrac{\alpha}{4\pi}+C-\dfrac{1}{2})+\mathcal{O}(\alpha^2)\\ 
&J_{+}^{(1)}(\alpha,0)= \dfrac{\pi^2}{6}- \dfrac{\alpha^2}{2}(\ln\dfrac{\alpha}{\pi}+C-\dfrac{1}{2})+\mathcal{O}(\alpha^2), \label{ac2}\\
&C\approx 0.577 \text{ (Euler Constant)}.
\end{cases}
\end{align}

The symmetry breaking scale is $Q=v_{20}$ and the minimum conditions for $V^{0K}_{eff}(v_2)$ as follows:
\begin{align}
V^{0K}_{eff}(v_{20})=0,\dfrac{\partial V^{0K}_{eff}(v_2)}{\partial v_2}(v_{20})=0,\dfrac{\partial^2 V^{0K}_{eff}(v_2)}{\partial v_2^2}(v_{20})=m_{H}^2(v_{20})+ m_{H_S}^2(v_{20}).\label{ac3}
\end{align}

Substituting Eqs.~\ref{ac1}, \ref{ac2} into $V_{eff}(v_2)= V^{0K}_{eff}(v_2)+V^{T}_{eff}(v_2)$ with the conditions of Eq.~\ref{ac3}. The one-loop effective potential of the $SU(2)_1\otimes SU(2)_2\otimes U(1)_Y \rightarrow SU(2)_L\otimes U(1)_Y$ phase transition which depends on the VEV $v_2$, can be rewritten as:
\begin{align}\label{clu}
V_{eff}(v_2)= \dfrac{\lambda_T}{4}v_2^4 -\theta Tv_2^3 + \gamma( T^2 - T_0^2) v_2^2,
\end{align} 
where
\begin{align*}
\lambda_T=&\dfrac{m_{H}^2(v_{20})+ m_{H_S}^2(v_{20})}{2v_{20}^2} \bigg\{ 1+ \dfrac{1}{8\pi^2 v_{20}^2 (m_{H}^2(v_{20})+ m_{H_S}^2(v_{20}))}  \bigg[ 
6m_{W'^\pm}^4(v_{20}) \ln \dfrac{bT^2}{m_{W'^\pm}^2(v_{20})}\\
+ &3m_{Z'}^4(v_{20}) \ln \dfrac{bT^2}{m_{Z'}^2(v_{20})} + m_{H}^4(v_{20}) \ln \dfrac{bT^2}{m_{H}^2(v_{20})} + m_{H_S}^4(v_{20}) \ln \dfrac{bT^2}{m_{H_S}^2(v_{20})}  \bigg]\bigg\}; b=49.4,\\
\theta=& \dfrac{1}{12\pi v_{20}^3}\bigg[ 6m_{W'^{\pm}}^3(v_{20}) +3m_{Z',}^3(v_{20})+ m_{H}^3(v_{20})+ m_{H_S}^3(v_{20}) \bigg],\\
\gamma =& \dfrac{1}{24 v_{20}^2 }\bigg[ 6m_{W'^{\pm}}^2(v_{20})  +3m_{Z'}^2(v_2)+ m_{H}^2(v_{20})+ m_{H_S}^2(v_{20})  \bigg],\\
T_0^2 =& \dfrac{1}{4\gamma}\bigg\{ m_{H}^2(v_{20})+ m_{H_S}^2(v_{20}) - \dfrac{1}{8\pi^2 v_{20}^2}\bigg[ 6m_{W'^{\pm}}^4(v_{20}) +3m_{Z'}^4(v_{20})+ m_{H}^4(v_{20}) + m_{H_S}^4(v_{20})   \bigg]\bigg\}.
\end{align*}

There are four unknown masses of Higgs bosons $H,H_S$ and two gauge bosons $W',Z'$. Notice that if $g_2\gg g'$, the mass terms of $W'$ and $Z'$ in this stage will be alike, i.e., their contribution in this phase transition are nearly equal. So we can consider the masses of these two gauge bosons as one variable and two Higgs bosons as another, in which the first variable is $m_{W'}(v_{20})= m_{Z'}(v_{20})=X$ and the second variable is $m_H(v_{20})=m_{H_S}(v_{20})=Y$. On the other hand, the masses of new gauge bosons have to be larger than $1.7$ TeV in order to satisfy the precision of $\rho-$parameter measurement Ref.~[\citen{221}], i.e., $X>1.7$ TeV.

\begin{figure}[http]	
	\centering
	\includegraphics[width=1\linewidth]{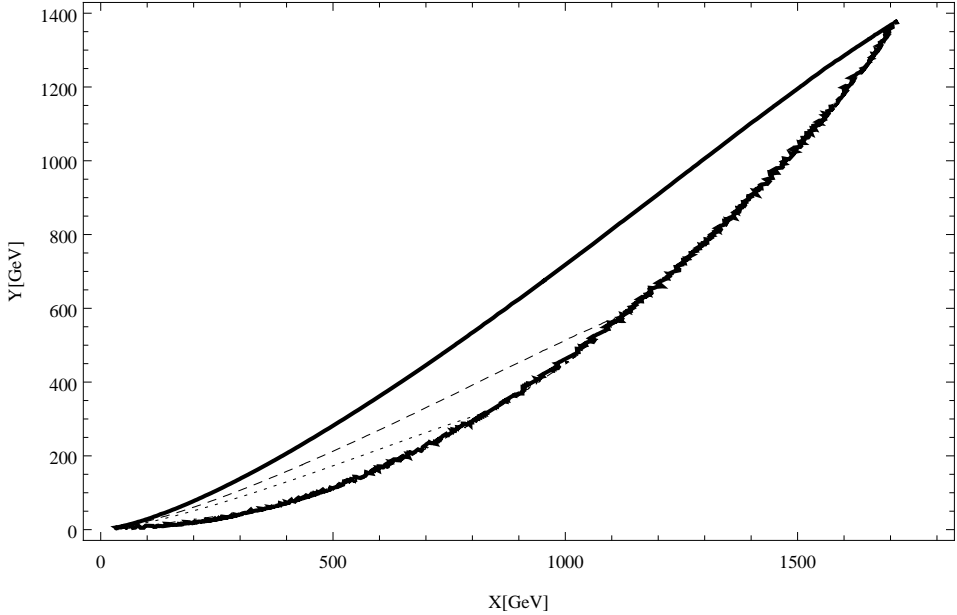}
	\caption{VEV $v_{20}=535$GeV. Thick contour $S=1$, dashed contour $S=1.5$, dotted contour $S=2$, dash-dotted contour $S_{max}=5$}
	\label{3S2-W-300}
\end{figure}

The first, we choose an arbitrary value of the symmetry breaking scale of this phase transition, e.g. $v_{20}=535$ GeV. After ploting $S$ as a function of $X,Y$ with the condition $S=\frac{2\theta}{\lambda_{T_C}}\geq 1$, we get Fig.~\ref{3S2-W-300} where we can see the upper limit of variable $X$. In Fig.~\ref{3S2-W-300}, the partly masses of $W'$ and $Z'$ are only about $1.7$ TeV. So we increase the scale and continue plotting. The last, we find that $v_{20}=535$ GeV is the least value that fits the $\rho-$ parameter condition and the range of the transition strength is $1\leq S < 5$.

We can see the range of unknown masses from Fig.~\ref{3S2-W-300} as:
\begin{align}
0\text{ GeV}< m_{W'}(v_{20})= m_{Z'}(v_{20})<1700 \text{ GeV}, \label{tss1}
\end{align}
and
\begin{align}
0\text{ GeV}< m_H(v_{20})=m_{H_S}(v_{20})<1370 \text{ GeV}, \label{tss2}
\end{align}
which means we have 
\begin{align}
0<\lambda_2<3.278, \qquad 0<g_2<3.06.
\end{align}

Then once again, back to the new gauge coupling, the authors of Ref.~[\citen{221}] have found that the constraint from the $\rho$ parameter becomes dominant when $g_2\ge 2$, which fits our result.

\begin{figure}[http]	
	\centering
	\includegraphics[width=1\linewidth]{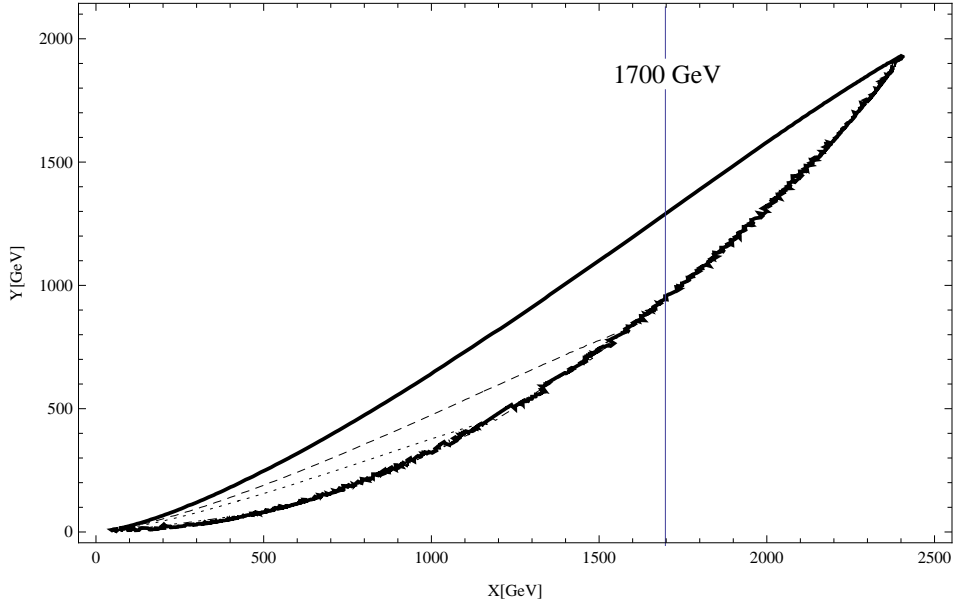}
	\caption{VEV $v_{20}=750$ GeV. Thick contour $S=1$, dashed contour $S=1.5$, dotted contour $S=2$, dash-dotted contour $S_{max}=4.8$}
	\label{650}
\end{figure}

Therefore, $v_{20}>535$ GeV, the mass of $Z'$ is larger than 1700 GeV. In Figure \ref{650}, we plot S with $v_ {20}=750$ GeV, the maximum mass of $Z'$ is about 2400 GeV. So the bigger $v_{20}$ is, the larger the mass of $Z'$ is.

\subsection{3rd phase transition $SU(2)_L\otimes U(1)_Y \rightarrow U(1)_Q$}
This phase transition involves a partly mass of new Higgs bosons $H_S$, a partly mass of new gauge boson $Z'$, with the masses of them are functions of $v$ as the 3rd column in table \ref{masss}. Importantly this phase involves $W^\pm$, $Z$, Higgs boson $h$ and top quark. This phase is SM-like but it has more new partilces.

In Table \ref{masss}, the partly mass of $H_S$ is $-\dfrac{\mu_{1S} v^2}{2\sqrt{2} v_S}$ which depend on $v, v_S$. This means that $H_S$ is involved in this phase. Because the dynamics variable is $v$ and we assume $v\ll v_S$ so $-\dfrac{\mu_{1S}}{2\sqrt{2} v_S}$ can be approximated as const. Therefore, the contribution of $H_S$ is considered for "an effective mass", i.e, $m_{H_s}^2(v)=const.v^2$.

The one loop effective potential of $SU(2)_L\otimes U(1)_Y \rightarrow U(1)_Q$ phase transition with the minimum conditions are:
\begin{align}
V^{0K}_{eff}(v_0)=0,\quad
\dfrac{\partial V^{0K}_{eff}(v)}{\partial v}(v_0)=0,\quad
\dfrac{\partial^2 V^{0K}_{eff}(v)}{\partial v^2}(v_0)=m_{h}^2(v_0).
\end{align}
The symmetry breaking scale is $v=246$ GeV. With the way as 2nd phase, the one-loop effective potential can be rewritten as:
\begin{align}
V_{eff}(v)= \dfrac{\lambda'_T}{4}v^4 -\theta 'Tv^3 + \gamma'( T^2 - T_0'^2) v^2,
\end{align} 
where
\begin{align*}
&\lambda'_T=\dfrac{m_{h}^2(v_0)}{2v_0^2} \bigg\{ 1+ \dfrac{1}{8\pi^2 v_0^2 m_{h}^2(v_0) } \bigg[ 
6m_{W^\pm}^4(v_0) \ln \dfrac{bT^2}{m_{W^\pm}^2(v_0)} +3m_{Z}^4(v_0) \ln \dfrac{bT^2}{m_{Z}^2(v_0)}\\
&+ 3m_{Z'}^4(v_0) \ln \dfrac{bT^2}{m_{Z'}^2(v_0)} + m_{h}^4(v_0) \ln \dfrac{bT^2}{m_{h}^2(v_0)} + m_{H_S}^4(v_0) \ln \dfrac{bT^2}{m_{H_S}^2(v_0)} -12m_{t}^4(v_0) \ln \dfrac{b_FT^2}{m_{t}^2(v_0)} \bigg]\bigg\};
b_F=3.12,\\
&\theta'= \dfrac{1}{12\pi v_0^3}\bigg[ 6m_{W^{\pm}}^3(v_0) +3m_{Z}^3(v_0) +3m_{Z'}^3(v_0)+ m_{h}^3(v_0)+ m_{H_S}^3(v_0) \bigg],\\
& \gamma' = \dfrac{1}{24 v_0^2 }\bigg[ 6m_{W^{\pm}}^2(v_0) +3m_{Z}^2(v_0) +3m_{Z'}^2(v_0)+ m_{h}^2(v_0)+ m_{H_S}^2(v_0) +6 m_t^2(v_0)  \bigg],\\
& T_0'^2 = \dfrac{1}{4\gamma'}\bigg\{ m_{h}^2(v_0)-\dfrac{1}{8\pi^2 v_0^2}\bigg[ 6m_{W^{\pm}}^4(v_0) +3m_{Z}^4(v_0) +3m_{Z'}^4(v_0)+ m_{h}^4(v_0)+ m_{H_S}^4(v_0) -12 m_t^4(v_0)  \bigg]\bigg\}.
\end{align*}
In this potential, we set the mass of SM-like Higgs boson $m_h(v_0)=125$ GeV then there are two unknown masses $m_{Z'}(v_0)$ and $m_{H_S}(v_0)$. Here, $\theta'$ has more distributions of $Z'$ and $H_S$ which do not appear in SM. The larger $\theta'$ is, the larger strength is. Therefore, the strength will be stronger than that of SM.    

\begin{figure}[!htp]
	\centering
	\includegraphics[width=1\linewidth]{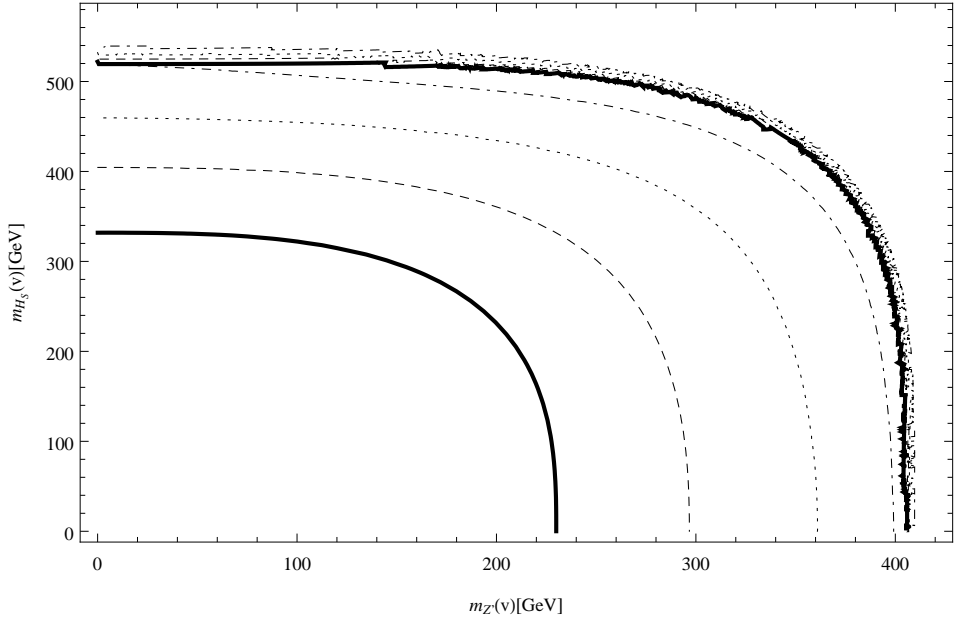}
	\caption{Thick contour $S=1$, dashed contour $S=1.2$, dotted contour $S=1.5$, dash-dotted contour $S_{max}=2.7$}\label{f1}
\end{figure}

In Fig.~\ref{f1}, $1\leq S=\frac{2\theta'}{\lambda'_{T_C}}<2.7$ can be found as the range of strength in this EWPT to be a first-oder transition.

As seen from Fig.~\ref{f1}, the range of unknown masses can be derived as below
\begin{align}
0<m_{Z'}(v_0) < 408\text{ GeV},\quad 95< m_{H_S}(v_0) < 524\text{ GeV}.\label{tss3}
\end{align}
Therefore, the mass of $H_S$ is at least about 100 GeV and we can obtain
\begin{align}
0<\frac14\frac{g'^4 v_0^2}{g_2^2 - g'^2}< (408\text{ GeV})^2, \quad (95\text{ GeV})^2<-\dfrac{\mu_{1S} v_0^2}{2\sqrt{2} v_{S}}< (524\text{ GeV})^2.\label{ts}
\end{align}

Because $v_S$ is a VEV at a different scale, it can not be a fixed value in this transition. Therefore, we can not get the range of $\mu_{1S}$ itself, which means we have to include $v_S$. From Eq. (\ref{ts}), we get
\begin{align}
g_2>0.41, \qquad -0.422>\dfrac{\mu_{1S}}{v_S}>-12.833.
\end{align}

The value of $g_2$ only has lower bound, no upper bound, consistent with the results in Ref.~[\citen{221}]. This result will be also combined with the remaining results in another phase transitions. The range of the new gauge coupling $g_2$ here also fits the $\rho$ parameter condition and does not lead to a Landau pole problem, which only occurs when $g_2$ reaches infinity. 

Note that this range, in Eq. (\ref{tss3}), is only part of $Z'$. Combining Eq.(\ref{tss1}), Eq.(\ref{tss2}) and Eq.(\ref{tss3}), we obtain a full mass range of $Z'$, $H_S$ and $W'$ as follows,

\begin{align}
0<m_{Z'} < \sqrt{408^2+1700^2}=1748.27 \text{ GeV},\\
0< m_{W'} < 1700\text{ GeV},\quad 0< m_{H} < 1370\text{ GeV},\\
95< m_{H_S} < \sqrt{524^2+1370^2}=1466.79 \text{ GeV}.\label{tss4}
\end{align}

\section{CONCLUSION AND OUTLOOKS}\label{sec8}

In this paper we have investigated the EWPT by using the high-temperature effective potential. The EWPT is strengthened by the new scalars to be the strongly first-order, the like SM phase transition strength is in the range $1 - 2.75$. Our results match the condition of $g_2>2$ in Ref.~[\citen{221}]. 

The expansion for a high-temperature effective potential will be better than $5\%$ if $\frac{m_{boson}}{T}< 2.2$ Ref.~[\citen{5percent}], where $m_{boson}$ is the relevant boson mass. The mass range in our calculations fit with this. 

The most research of EWPT are in the Landau gauge. A gauge is not importantly contributions in the EWPT, it mean that we can ignore the role of Goldstone bosons, as researching in Refs.~[\citen{1101.4665,Arefe}]. Therefore, our research about EWPT in the Landau gauge is also sufficient.

The damping effect is in the thermal self-energy term $(\Sigma_{ij}(T)\phi_i\phi_j$ and $\Pi^{ab}(T)A^a_0A^b_0$, i.e., $V^B_{ring}$ in Ref.~[\citen{1101.4665}]. This effect which is from the ring loop distribution, still is very small. It was approximated $g^2T^2/m^2$ ($g$ is the coupling constant of $SU(2)$, $m$ is mass of boson), $m\sim 100$ GeV, $g\sim 10^{-1}$ so $g^2/m^2 \sim 10^{-6}$. If we add this distribution to the effective potential, the square terms will give a small change  only. Therefore, this distribution does not change the strength of EWPT or, in other words, it  is not the origin of EWPT. 

In this model, the mixing between normal quark and exotic quarks (VLQ) is shown through the $\theta^q_L$ ($\tan\theta^q_L=\frac{m_qm_{qQ}}{m^2_F}$, $m_q$ is the masses of normal quarks, $m_F$ is the masses of VLQs, $m_{qQ} << m_F$; $\tan\theta^q_R=\frac{m_{qQ}}{m_F}$, in detail to Ref.~[\citen{221}]), the masses of VLQs are very larger than of normal quarks so that $\theta^q_L \sim 0$. Therefore, FCNC can not appear at the tree level in this model. The mass of Z' is larger than 1.7 TeV, this is not enough for turning on FCNC with normal quarks. The Current empirical data also shows that the mass of Z' is about some TeVs and FCNC with VLQs can exist, at the TeV energy scale.

$H_S$ is a complex case in this model, because its mass are intertwined between the VEVs. This complicates the separation of phase in subsequent calculations. Also follow that the second VEV has not an upper bound so needing more constraints in the decay channels of Higgs. Therefore we will introduce a Higgs potential correction to determine clearly the mass of $H_S$ and $v_{20}$.   

Furthermore, the sphaleron is an important process in baryogenesis so we will continue to calculate and test the sphaleron solution in this model with the Cosmotransition code Ref.~[\citen{cosmotran}]. This code used a Bessel function for $v(r)$ but it is not flexible in changing the value of wall.

This work could serve as the basis for the calculation of cross section of the decay Higgs to photons when connected to the data of LHC or Particle Data Group.

\section*{ACKNOWLEDGMENTS}
This research is funded by Vietnam National University Ho Chi Minh City (VNU-HCM) under grant number C2017-18-12.

\end{document}